\documentstyle[pra,aps,twocolumn]{revtex}

\def\H{{\cal H}} 
\def\tr{{\rm tr}}
\def\kb#1{\vert#1\rangle\langle#1\vert} 
\def\Cx{\hbox{\kern.3em\vrule height 1.55ex depth -.1ex width .2pt\kern-.3em\rm C}}

\title{A counterexample to a conjectured entanglement inequality}
 \author{K.~G.~H. Vollbrecht\thanks{Electronic Mail:
\tt{k.vollbrecht@tu-bs.de}}{{}\quad and\ } 
  R.~F. Werner\thanks{Electronic Mail: \tt{R.Werner@tu-bs.de}}
  \\[1ex]
  {\small Institut f{\"u}r Mathematische Physik, TU Braunschweig,
   Mendelssohnstr.3, 38106 Braunschweig, Germany.}}

\begin{document}
\maketitle

\begin{abstract}\noindent
We give an explicit counterexample to an entanglement inequality
suggested in a recent paper [quant-ph/0005126] by Benatti and
Narnhofer. The inequality would have had far-reaching
consequences, including the additivity of the entanglement of
formation.
\end{abstract}

\pacs{}

The problem of additivity of entanglement of formation
\cite{c:eof} is one of the fundamental open issues in the theory
of entanglement. In a recent paper \cite{c:benar} Fabio Benatti
and Heide Narnhofer bring to bear on this problem their intuition
gained from a completely different enterprise (the study of
quantum dynamical entropy), using the equivalence between the
entanglement of formation and a quantity called the ``entropy of a
subalgebra'' with respect to a state (a connection noted earlier
by Uhlmann \cite{c:uhl} and others). They achieve some interesting partial 
results supporting the additivity conjecture\footnote{We only 
mention in passing an omission in the formulation of  
Proposition~1 in \cite{c:benar}: the unitary $U$ must be assumed 
to factorize. }. 

In this brief note we will concentrate on an exciting prospect 
coming up in their paper as inequality (12) (restated below as 
equation~(\ref{ineq})). Benatti and Narnhofer apparently found it 
as an inequality one would just love to have for a simple proof of 
additivity, but seem undecided as to its validity. Given the 
simplicity and generality of the inequality, probably many more 
results would follow from it. It therefore seemed necessary to us 
to decide the issue quickly. Unfortunately, (if one can ever say 
that of a mathematical statement) the inequality turned out to be 
false in general. 

The inequality in question refers to a vector $\Psi$ in a fourfold
tensor product $\Psi\in\H_1\otimes\H_2\otimes\H_3\otimes\H_4$, of
which the factors $\H_1$ and $\H_3$ are associated with one party
(say Alice) and factors $\H_2$ and $\H_4$ are associated with
another, Bob. We can write $\Psi$ in Schmidt form as
\begin{equation}\label{Schmidt}
  \Psi=\sum_\alpha\sqrt{\lambda_\alpha}\;
       \Phi^{12}_\alpha\otimes\Phi^{34}_\alpha
\end{equation}
with respect to another split $12\vert34$, of the system, which is
the split according to which additivity would be investigated. Of
course, by definition of the Schmidt decomposition, we have
$\sum_\alpha\lambda_\alpha=1$, and the vectors
$\Phi^{12}_\alpha\in\H_1\otimes\H_2$ and
$\Phi^{34}_\alpha\in\H_3\otimes\H_4$ each form an orthonormal set.
Let us denote by $S(\rho)$ the von Neumann entropy of a density
operator $\rho$, and by $\tr_i(A)$ (resp. $\tr_{ij}(A)$) the
partial trace of the operator $A$ with respect to the Hilbert
space $\H_i$ (resp. $\H_i\otimes\H_j$), assumed to be a tensor
factor of the space on which $A$ lives. Then the inequality
suggested by Benatti and Narnhofer is
\begin{eqnarray}\label{ineq}
  S\Bigl(\tr_{24}(\kb\Psi)\Bigr)
  &\geq&
    \sum_\alpha\lambda_\alpha\Bigl\lbrace
      S\bigl(\tr_{2}(\kb{\Phi^{12}_\alpha})\bigr)+
\nonumber\\ &&\qquad+
      S\bigl(\tr_{4}(\kb{\Phi^{34}_\alpha})\bigr)
      \Bigr\rbrace\;.
\end{eqnarray}

A superficial random numerical test might come out in favor of
(\ref{ineq}). A counterexample is found, however, with all Hilbert
spaces $\H_i=\Cx^d$, $d$ arbitrary, namely
\begin{equation}\label{psiex}
  \Psi=\frac1d\sum_{i,k=1}^d e_i\otimes e_k\otimes e_i\otimes e_k
      =\frac1d\sum_{\alpha=1}^{d^2} \Phi_\alpha\otimes
      \overline{\Phi_\alpha}\;,
\end{equation}
where the vectors $e_i$ form an orthonormal basis of $\Cx^d$. Thus
$\Psi$ is the tensor product of the vector
$d^{-1/2}\sum_ie_i\otimes e_i$ for Alice and the same vector for
Bob, making the left hand side of inequality (\ref{ineq}) zero.
$\Psi$ is also maximally entangled for the split $12\vert34$. One
Schmidt decomposition is into the vectors $e_i\otimes e_k$ (which
would make the right hand side of (\ref{ineq}) zero, too). But in
the maximally entangled case the Schmidt decomposition is highly
non-unique, and we may also choose another decomposition using
entangled vectors $\Phi_\alpha$ and their complex conjugates
$\overline{\Phi_\alpha}$ with respect to the basis $e_i$. We may
even take them maximally entangled \cite{c:2qb}, whence the right
hand side of (\ref{ineq}) becomes $2\log d$. By deforming the
coefficients in the latter decomposition slightly, we also get
examples with unique Schmidt decomposition such that
LHS(\ref{ineq})$\approx0$ and RHS(\ref{ineq})$\approx2\log d$.


\begin{references}
\bibitem{c:eof}C.H. Bennett et al, {\it Phys.Rev. A \bf54}
(1996)3824, quant-ph/9604024

\bibitem{c:benar}F. Benatti and H. Narnhofer: ``On the additivity
of the entanglement of formation'', quant-ph/0005126.

\bibitem{c:uhl}A. Uhlmann:``Entropy and optimal decompositions of 
states relative to a maximal commutative subalgebra'', quant-ph/9704017 

\bibitem{c:2qb}K.G.H. Vollbrecht and R.F. Werner: Why two qubits are special,
quant-ph/9910064.
\end{references}
\end{document}